\documentclass[aps,prl,amsmath,amssymb,nofootinbib,tightenlines,floatfix,superscriptaddress]{revtex4}
\usepackage[dvips]{graphics}
\usepackage{epsfig}
\usepackage{latexsym,amssymb}
\newcommand{\lsim}{\buildrel < \over {_\sim}}

\newcommand{\be}{\begin{equation}}
\newcommand{\ee}{\end{equation}}
\newcommand{\bea}{\begin{eqnarray}}
\newcommand{\eea}{\end{eqnarray}}
\begin{document}
\rightline{SLAC-PUB-16238}

\title{Comment on 
``New Limits on Intrinsic Charm in the Nucleon from Global Analysis of 
Parton Distributions''}

\author{Stanley J. Brodsky}
\affiliation{SLAC National Accelerator Laboratory, Stanford University, Stanford, CA 94309} 
\author{Susan Gardner}
\affiliation{Department of Physics and Astronomy, University of Kentucky, Lexington, KY 40506-0055}
\begin{abstract} 
A Comment on the Letter by 
P.~Jimenez-Delgado, T.~J.~Hobbs, 
J.~T.~Londergan, and W.~Melnitchouk, 
Phys. Rev. Lett. {\bf 114}, 082002 (2015). 
\end{abstract}

\maketitle 

Intrinsic heavy quarks in hadrons emerge from the 
non-perturbative structure of a hadron bound state~\cite{Brodsky:1980pb} 
and are a rigorous prediction 
of QCD~\cite{Brodsky:1984nx,Franz:2000ee}. 
Lattice QCD calculations
indicate significant intrinsic 
charm and strangeness probabilities~\cite{Freeman:2012ry,Gong:2013vja}.
Since the light-front momentum distribution of the 
Fock states is maximal at equal rapidity, 
intrinsic heavy quarks carry significant fractions of the hadron
momentum. 
The presence of Fock states 
with intrinsic strange, charm, or bottom quarks 
in hadrons lead to an array of 
novel physics phenomena~\cite{Brodsky:2014bla,Lyonnet:2015dca,Brodsky:2015fna}. 
Accurate determinations of the heavy-quark distributions 
in the proton are needed to interpret Tevatron and 
LHC measurements as probes of physics beyond the 
Standard Model~\cite{Dulat:2013hea,Brambilla:2014jmp}. 
Determinations~\cite{Harris:1995jx,Pumplin:2007wg,Dulat:2013hea} of the 
momentum fraction carried by intrinsic charm quarks in the proton typically
limit $\langle x \rangle_{\rm IC} \sim {\cal O}(1\%)$ at 90\% CL, 
consistent with the EMC analysis 
of their charm structure function data~\cite{Aubert:1982tt} 
and the large rate for high-$p_T$ 
$\bar p p \to c \gamma X$ reactions at the Tevatron~\cite{Mesropian:2014kfa};
however, a precise determination of $\langle x \rangle_{\rm IC}$ has 
proven elusive. 

The recent letter 
by P.~Jimenez-Delgado et al.~(JDHLM)~\cite{Jimenez-Delgado:2014zga} reports
$\langle x\rangle_{\rm IC} =(0.15\pm 0.09)\%$. The authors
include low-energy data 
from the 
$e d\, (p) \to e^\prime X$ 
SLAC experiment~\cite{Whitlow:1991uw}
in their  global fit. They find that this data set 
places strong
constraints on intrinsic charm, although, by their count, only 157 of 
1021 data points have $W^2$ in excess of the charm hadronic
threshold: $W^2_{\rm th}\approx 16\,{\rm GeV}^2$. 
The SLAC measurements of 
$e d \,(p)  \to e^\prime X$ 
have an overall normalization 
(systematic) error of $\pm~1.7\, (2.1) \% $, and
a relative normalization error of 
typically $\pm 1.1\%$~\cite{Whitlow:1991uw}.  
The SLAC data points in the germane 
regime of $W^2 \gtrsim 16~{\rm GeV}^2,\, x >0.1$ 
have even larger statistical uncertainties. 

It is clearly challenging to identify the contribution from charm quarks
to the inclusive 
structure function 
if only the scattered electron is detected. 
In addition to the valence and sea-quark distributions, 
there are other contributions to the inclusive cross section 
which need to be determined to high accuracy in order to discern the 
intrinsic charm component at the level claimed by JDHLM; this includes 
higher-twist corrections at high $x$, 
the strange and bottom quark contributions, as well as 
the accurate implementation of the suppression of charm 
production at threshold and nuclear target effects. 
Their analysis depends on 
uncertain parameters and 
theory assumptions.
For example, JDHLM model higher-twist effects as an isospin-independent, 
phenomenological multiplicative factor on top of a leading-twist
structure function with target-mass corrections~\cite{Owens:2012bv}.   
Their higher-twist model does not include enhancements
at $x \sim 1$ resulting from hard subprocesses such as 
$e [qq] \to e^\prime q q $ where 
multiple quarks interact~\cite{Berger:1979du,Brodsky:1982qn}. 
Such processes depend strongly on the diquark charges, making them 
quark flavor (isospin) sensitive, and they contribute
in the same large $x$ domain as the charm signal. 
In addition, 
the target mass-corrected structure functions used by JDHLM remain nonzero 
as $x\to 1$~\cite{Owens:2012bv} which is problematic, and they neglect  intrinsic 
strange and beauty quark contributions~\cite{Ellis:2002zv,Lyonnet:2015dca}.

JDHLM assess their PDF errors using a tolerance criteria of 
$\Delta \chi^2=1$ at 1$\sigma$; however, 
the actual value of $\Delta \chi^2$  
depends on the number of parameters to be simultaneously 
determined in the fit --- their assessment of a single parameter error 
requires that the other parameters be fixed 
at their values at the global $\chi^2$ minimum~\cite{pdg}.
JDHLM report  
$\langle x\rangle_{\rm IC} =(0.15\pm 0.09)\%$~\cite{Jimenez-Delgado:2014zga} 
corresponding to $\Delta \chi^2=1$ at 1$\sigma$ (68\% CL)
and also $\langle x\rangle_{\rm IC} \lsim 0.5\%$ at 4$\sigma$. 
In order to set a $4\sigma$ limit,  all of the other parameters
must be varied so as to yield a minimum $\chi^2$ as the parameter of interest
is changed~\cite{avni,ckmfitter,Liu:2007yi,Gardner:2013aya}. 
We note Refs.~\cite{Gao:2013xoa,Jimenez-Delgado:2014twa}, e.g., contain
25 PDF parameters in leading twist, and Ref.~\cite{Jimenez-Delgado:2014twa}
contains 12 more higher twist parameters. 
Since one would expect nontrivial
correlations and near degeneracies in a many-parameter fit, 
the apparent agreement of the $4 \times 1\sigma$ assessment
with the $4\sigma$ limit suggests
that the other parameters were not properly varied 
as $\langle x\rangle_{\rm IC}$ was changed,
making the reported limit too 
severe. 

It is clear that 
the SLAC single-arm measurements 
cannot unambiguously identify an intrinsic charm contribution to the nucleon 
structure function even at the $1\%$ level 
because of statistical and systematic uncertainties, 
both experimental and theoretical.
JDHLM have excluded the EMC measurements 
of the charm structure function~\cite{Aubert:1982tt}, citing  a ``goodness of 
fit" criterion. 
However, statistical criteria alone cannot exclude data
sets. The fit to the EMC iron target data would be improved by including 
the QCD nuclear effects 
described in Refs.~\cite{deFlorian:2011fp,deFlorian:2012qw}. 

In summary, JDHLM claim that the momentum fraction carried by 
intrinsic charm  quarks in the proton is 
$\langle x\rangle_{\rm IC} =(0.15\pm 0.09)\%$; 
we do not find this conclusion warranted. 

\begin{acknowledgments}
We thank 
B.~Plaster, A.~Deur, P.~Hoyer, Ali~Khorramian, C.~Lorc{\'e}, G.~Lykasov, J..~Pumplin,
and R.~Vogt for helpful discussions. 
We acknowledge support from 
the U.S. Department of Energy 
under contracts DE--AC02--76SF00515 and DE--FG02--96ER40989. 
\end{acknowledgments}

\end{document}